\documentclass{appolb}

\usepackage{graphicx}
\usepackage{amsmath}

\newcommand{\ltapprox}{\raisebox{-0.5ex}{$\,\stackrel{<}{\scriptstyle\sim}\,$}}

\begin{document}


\title{Masses of mesons with charm valence quarks from $2+1+1$ flavor twisted
mass lattice QCD\thanks{Presented at ``Excited QCD 2013'', Bjelasnica Mountain, Sarajevo.}}

\author{
Martin Kalinowski, Marc Wagner
\address{
Goethe-Universit\"at Frankfurt am Main, Institut f\"ur Theoretische Physik, Max-von-Laue-Stra{\ss}e 1, D-60438 Frankfurt am Main, Germany \\
European Twisted Mass Collaboration
}
}

\maketitle

\begin{abstract}
We present preliminary results of an ongoing lattice QCD computation of the spectrum of $D$ mesons and $D_s$ mesons and of charmonium using 2+1+1 flavors of twisted mass sea and valence quarks.
\end{abstract}

% 12.38.Gc   Lattice QCD calculations (see also 11.15.Ha Lattice gauge theory)
% 14.40.Lb   Charmed mesons (|C|>0, B=0) 

% ********************
% ********************
% ********************

\section{Introduction}

There is considerable interest in the spectrum of $D$ and $D_s$ mesons and of charmonium both theoretically and experimentally.

On the theory side first principles calculations are usually lattice QCD computations (for recent work cf.\ e.g.\ \cite{Mohler:2011ke,Namekawa:2011wt,Liu:2012ze,Dowdall:2012ab,Bali:2012ua,Moir:2013ub}). In the last couple of years a lot of progress has been made, allowing the determination of hadron masses like the aforementioned mesons with rather high precision. For example 2+1 or even 2+1+1 flavors of dynamical quark are often used as well as small lattice spacings and improved discretizations, to keep discretization errors (in particular those, associated with the heavy charm quarks) under control. Some groups have even started to determine the resonance parameters of certain mesons from the spectrum of two-particle scattering states in finite spatial volumes (cf.\ e.g.\ \cite{Mohler:2012na}).

Experimentally a large number of $D$, $D_s$ and charmonium states has been measured and additional and/or more precise results are expected in the near future both from existing facilities and facilities currently under construction, like the PANDA experiment at FAIR. Even though these experimental results have been extremely helpful, to improve our understanding of QCD, they also brought up new and yet unanswered questions. For example the positive parity mesons $D_{s0}^*$ and $D_{s1}$ are unexpectedly light, which is at the moment not satisfactorily understood and also quite often not reproduced by lattice QCD computations or model calculations.

Moreover, performing a precise computation of certain meson masses is often the first step for many lattice projects not primarily concerned with spectroscopy. As an example one could mention the semileptonic decay of $B$ and $B^*$ mesons into positive parity $D$ mesons \cite{Bigi:2007qp}, whose masses and operator contents are an essential ingredient for any corresponding lattice computation.

This is a status report about an ongoing lattice QCD project concerned with the computation of the spectrum of mesons with at least one charm valence quark. We present preliminary results for $D$ mesons, for $D_s$ mesons and for charmonium states with total angular momentum $J = 0, 1$ and parity $P = -, +$. Parts of this work have already been published \cite{Kalinowski:2012re}.

% ********************
% ********************
% ********************

\section{Simulation and analysis setup}

We use gauge link configurations with 2+1+1 dynamical quark flavors generated by the European Twisted Mass Collaboration (ETMC) \cite{Baron:2008xa,Baron:2009zq,Baron:2010bv,Baron:2011sf,Baron:2010th,Baron:2010vp}. Until now we have considered two ensembles (around 600 gauge link configurations per ensemble) with (unphysically heavy) values for the light $u/d$ quark mass corresponding to $m_\pi \approx 325 \, \textrm{MeV}, 457 \, \textrm{MeV}$ (lattice sizes $(L/a)^3 \times T/a = 32^3 \times 64, 24^3 \times 48$). Our results are obtained at a single lattice spacing $a \approx 0.086 \, \textrm{fm}$. Consequently, a continuum extrapolation has not yet been performed.

Meson masses are determined by computing and studying temporal correlation matrices of suitably chosen meson creation operators $\mathcal{O}_j$. At the moment we exclusively consider quark antiquark operators. The quark and the antiquark are combined in spin space via $\gamma$ matrices and in color and position space via gauge links (discretized covariant derivatives) such that the corresponding trial states $\mathcal{O}_j | \Omega \rangle$ ($| \Omega \rangle$ denotes the vacuum) are gauge invariant and have defined total angular momentum and parity. Moreover, APE and Gaussian smearing is used, to optimize the overlap of the trial states $\mathcal{O}_j | \Omega \rangle$ to the low lying mesonic states of interest. More details regarding the construction of meson creation operators in twisted mass lattice QCD can be found e.g.\ in \cite{Jansen:2008si}. We plan to discuss these operators, their structure and their quantum numbers in detail in an upcoming publication. For the computation of the corresponding 
correlation matrices $\langle \mathcal{O}_j^\dagger(t) \mathcal{O}(0) \rangle$ we resort to the one-end trick (cf.\ e.g.\ \cite{Boucaud:2008xu}). Meson masses are then determined from plateaux values of corresponding effective masses, which we obtain by solving generalized eigenvalue problems (cf.\ e.g.\ \cite{Blossier:2009kd}). Disconnected diagrams appearing in charmonium correlators are currently neglected.

For both the valence strange and charm quarks we use degenerate twisted mass doublets, i.e.\ a different discretization as for the corresponding sea quarks. We do this, to avoid mixing of strange and charm quarks, which inevitably takes place in a unitary setup, and which is particularly problematic for hadrons containing charm quarks \cite{Baron:2010th,Baron:2010vp}. The degenerate valence doublets allow two realizations for strange as well as for charm quarks, either with a twisted mass term $+i \mu_{s,c} \gamma_5$ or $-i \mu_{s,c} \gamma_5$. For a quark antiquark meson creation operator the sign combinations $(+,-)$ and $(-,+)$ for the quark $q$ and the antiquark $\bar{q}$ are related by symmetry, i.e.\ the corresponding correlators are identical. These correlators differ, however, from their counterparts with sign combinations $(+,+)$ and $(-,-)$, due to different discretization errors. In section~\ref{SEC001} we will show for each computed meson mass both the $(+,-) \equiv (-,+)$ and the $(+,+) \equiv (-
,-)$ result. The differences are $\mathcal{O}(a^2)$ due to automatic $\mathcal{O}(a)$ improvement inherent to the twisted mass formulation. These mass differences give a first impression regarding the magnitude of discretization errors at our currently used lattice spacing.

Using $(+,-) \equiv (-,+)$ correlators we have tuned the bare valence strange and charm quark masses $\mu_s$ and $\mu_c$ to reproduce the physical values of $2 m_K^2 - m_\pi^2$ and $m_D$, quantities, which strongly depend on $\mu_s$ and $\mu_c$, but which are essentially independent of the light $u/d$ quark mass.

% ********************
% ********************
% ********************

\section{\label{SEC001}Numerical results}

\begin{figure}[t]
\begin{center}
\includegraphics[width=8.5cm]{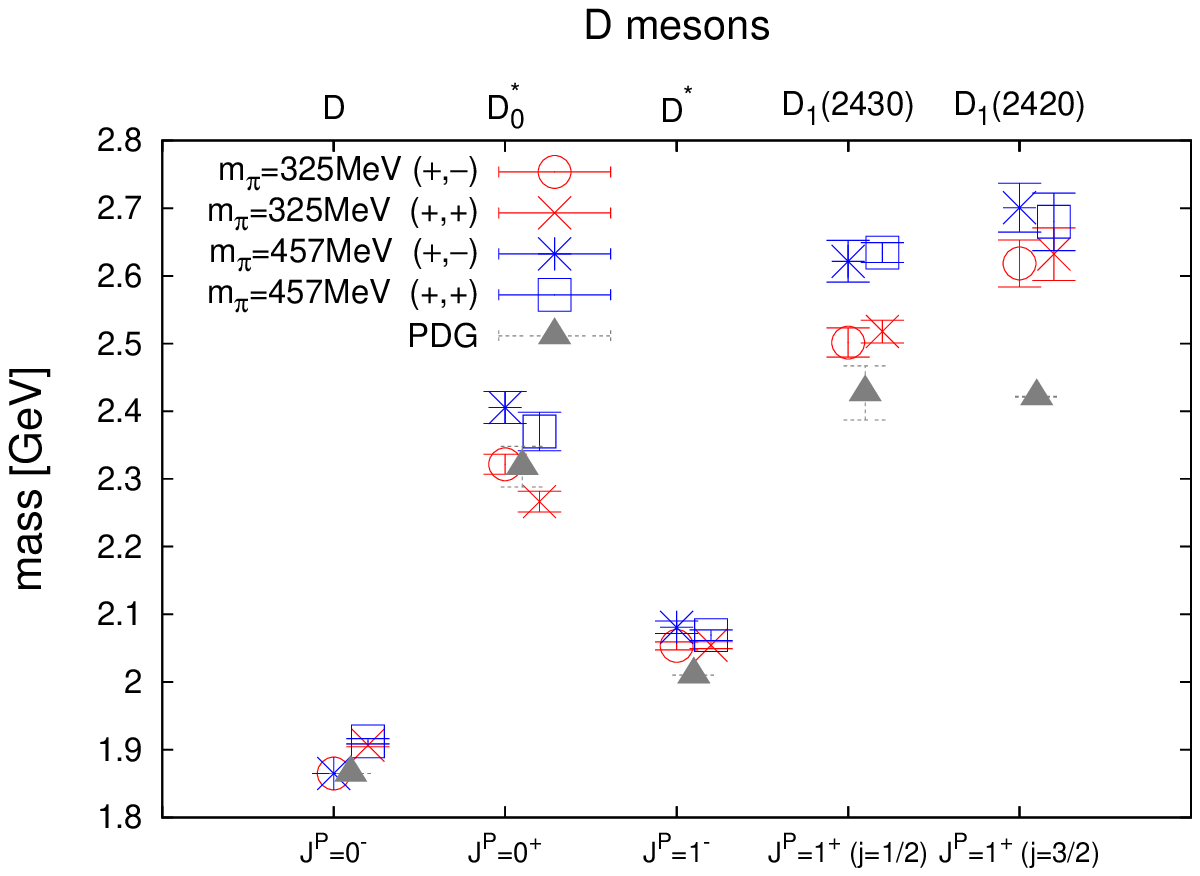} \\
\includegraphics[width=8.5cm]{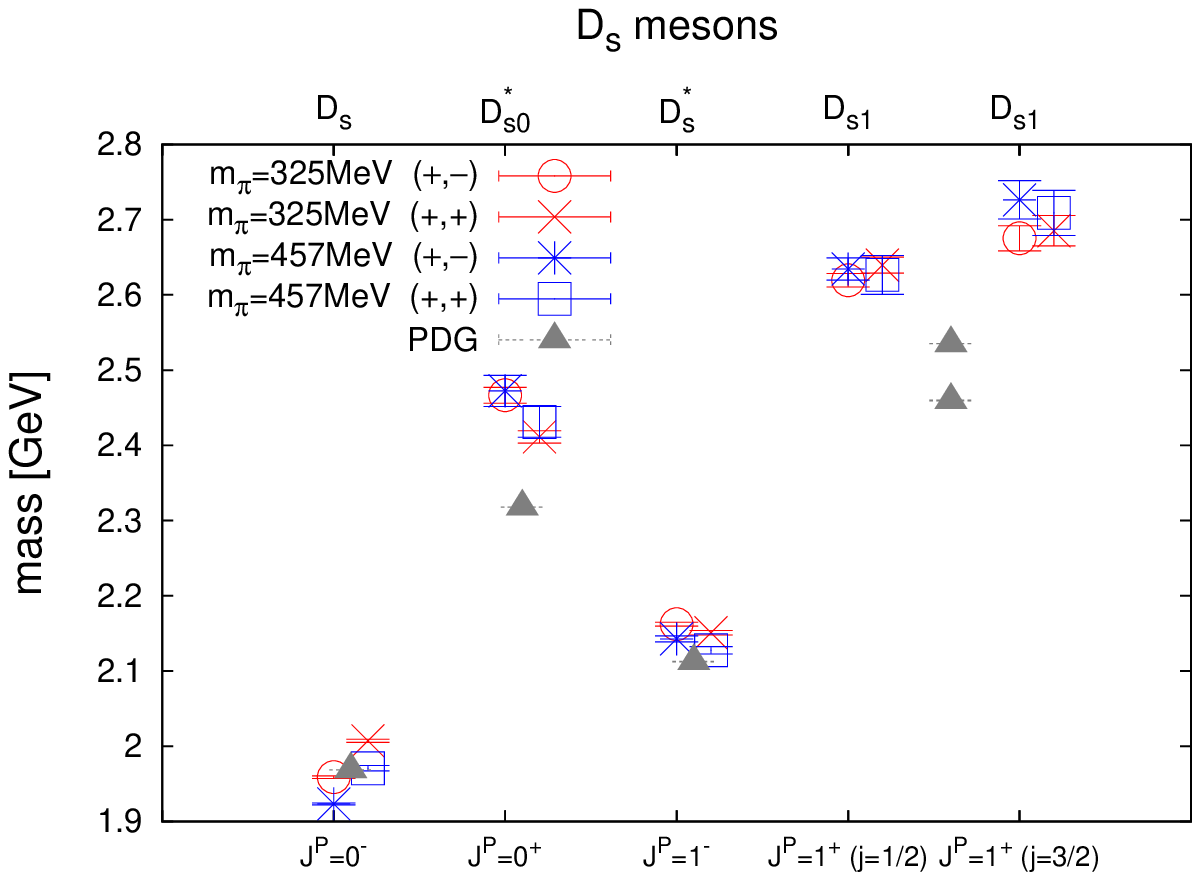}
\end{center}
\vspace{-0.5cm}
\caption{\label{FIG001}The $D$ and $D_s$ meson spectrum.}
\end{figure}

In Fig.~\ref{FIG001} we present our results for the $D$ and $D_s$ meson spectrum. For every state we show five data points:
\\ \phantom{XXX} \textit{Red circles and crosses}: \\ \phantom{XXXXXX} lattice results at $m_\pi \approx 325 \, \textrm{MeV}$, twisted mass sign combinations \\ \phantom{XXXXXX} $(+,-) \equiv (-,+)$ and $(+,+) \equiv (-,-)$, respectively.
\\ \phantom{XXX} \textit{Blue stars and boxes}: \\ \phantom{XXXXXX} lattice results at $m_\pi \approx 457 \, \textrm{MeV}$, twisted mass sign combinations \\ \phantom{XXXXXX} $(+,-) \equiv (-,+)$ and $(+,+) \equiv (-,-)$, respectively.
\\ \phantom{XXX} \textit{Gray triangles}: \\ \phantom{XXXXXX} experimental result from the PDG \cite{PDG}.
\\ The differences between sign combinations $(+,-) \equiv (-,+)$ and \\ $(+,+) \equiv (-,-)$, which are $\ltapprox 3 \%$, indicate the magnitude of discretization errors at our currently used lattice spacing $a \approx 0.086 \, \textrm{fm}$.

% *****

% *****

While for the negative parity states lattice and experimental results agree rather well, there is a clear discrepancy in particular for the positive parity $D_s$ states $D_{s0}^*$ and $D_{s1}$. Similar findings have been reported in other lattice studies, e.g.\ \cite{Mohler:2011ke,Moir:2013ub}, and in phenomenological model calculations, e.g.\ \cite{Ebert:2009ua}. This discrepancy might be an indication that these states are not predominantly $q \bar{q}$ states, but e.g.\ rather four quark states of molecular or tetraquark type. We plan to investigate this possibility within our setup in the near future. The necessary techniques have already been developed and recently been applied to light scalar mesons \cite{Alexandrou:2012rm}.

Another challenging, but important task is the separation of the two $J=1^+$ states, $D_1(2420), D_1(2430)$ and $D_{s1}(2460), D_{s1}(2535)$, respectively. In the limit of a static charm quark one of these states has light cloud angular momentum $j = 1/2$, while the other has $j = 3/2$. To assign corresponding approximative $j$ quantum numbers, when using charm quarks of finite mass, is e.g.\ important, when studying the decay of a $B$ or $B^*$ meson to one of the positive parity $D^{**}$ mesons (which include the mentioned $D_1(2420)$ and $D_1(2430)$ states) in a fully dynamical setup (cf.\ e.g.\ \cite{Blossier:2009vy,Blossier:2009eg} for a recent lattice computation in the static limit). The correct identification of $j = 1/2$ and $j = 3/2$ states can be achieved by studying the eigenvectors obtained during the analysis of correlation matrices; the largest eigenvector components point out the dominating operators, which, after a Clebsch-Gordan decomposition into light and heavy angular momentum 
contributions, can be classified according to $j = 1/2$ or $j = 3/2$.

\begin{figure}[h]
\begin{center}
\includegraphics[width=8.5cm]{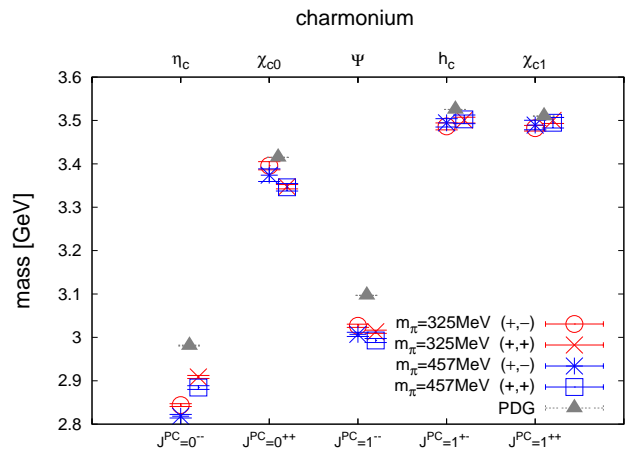}
\end{center}
\vspace{-0.5cm}
\caption{\label{FIG002}The charmonium spectrum.}
\end{figure}

In Fig.~\ref{FIG002} we present our results for the charmonium spectrum. Because of the two rather heavy valence quarks, we expect considerably larger discretization errors as for the corresponding $D$ or $D_s$ meson states. The differences between lattice and experimental results are most prominent for the negative parity charmonium states (around $5 \%$). We plan to explore in one of our next steps, whether discretization errors account for these differences by performing similar computations on ensembles with finer lattice spacings and by studying the continuum limit.

% *****

% *****

% ********************
% ********************
% ********************

\section*{Acknowledgments}

We thank Christian Wiese for discussions. M.K.\ and M.W.\ acknowledge support by the Emmy Noether Programme of the DFG (German Research Foundation), grant WA 3000/1-1 and by Helmholtz Graduate School HGS-HIRe for FAIR. This work was supported in part by the Helmholtz International Center for FAIR within the framework of the LOEWE program launched by the State of Hesse.

% ********************
% ********************
% ********************


\begin{thebibliography}{99}
\bibliographystyle{ieeetr}

\bibitem{Mohler:2011ke} 
  D.~Mohler and R.~M.~Woloshyn,
  % ``$D$ and $D_s$ meson spectroscopy,''
  Phys.\ Rev.\ D {\bf 84}, 054505 (2011)
  [arXiv:1103.5506 [hep-lat]].
  %%CITATION = ARXIV:1103.5506;%%	

\bibitem{Namekawa:2011wt}
  Y.~Namekawa {\it et al.} [PACS-CS Collaboration],
  %``Charm quark system at the physical point of 2+1 flavor lattice QCD,''
  Phys.\ Rev.\ D {\bf 84} (2011) 074505
  [arXiv:1104.4600 [hep-lat]].
  %%CITATION = ARXIV:1104.4600;%%

\bibitem{Liu:2012ze} 
  L.~Liu {\it et al.} [Hadron Spectrum Collaboration],
  %``Excited and exotic charmonium spectroscopy from lattice QCD,''
  JHEP {\bf 1207}, 126 (2012)
  [arXiv:1204.5425 [hep-ph]].
  %%CITATION = ARXIV:1204.5425;%%

\bibitem{Dowdall:2012ab}
  R.~J.~Dowdall, C.~T.~H.~Davies, T.~C.~Hammant and R.~R.~Horgan,
  %``Precise heavy-light meson masses and hyperfine splittings from lattice QCD including charm quarks in the sea,''
  Phys.\ Rev.\ D {\bf 86} (2012) 094510
  [arXiv:1207.5149 [hep-lat]].
  %%CITATION = ARXIV:1207.5149;%%

\bibitem{Bali:2012ua}
  G.~Bali, S.~Collins and P.~Perez-Rubio,
  %``Charmed hadron spectroscopy on the lattice for $N_f=2+1$ flavours,''
  J.\ Phys.\ Conf.\ Ser.\  {\bf 426} (2013) 012017
  [arXiv:1212.0565 [hep-lat]].
  %%CITATION = ARXIV:1212.0565;%%

\bibitem{Moir:2013ub} 
  G.~Moir, M.~Peardon, S.~M.~Ryan, C.~E.~Thomas and L.~Liu,
  %``Excited spectroscopy of charmed mesons from lattice QCD,''
  arXiv:1301.7670 [hep-ph].
  %%CITATION = ARXIV:1301.7670;%%

\bibitem{Mohler:2012na} 
  D.~Mohler, S.~Prelovsek and R.~M.~Woloshyn,
  %``D Pi scattering and D meson resonances from lattice QCD,''
  Phys.\ Rev.\ D {\bf 87}, 034501 (2013)
  [arXiv:1208.4059 [hep-lat]].
  %%CITATION = ARXIV:1208.4059;%%


\bibitem{Bigi:2007qp} 
  I.~I.~Bigi {\it et al.},
  %``Memorino on the `1/2 vs. 3/2 Puzzle' in anti-B ---> l anti-nu X(c): A Year Later and a Bit Wiser,''
  Eur.\ Phys.\ J.\ C {\bf 52}, 975 (2007)
  [arXiv:0708.1621 [hep-ph]].
  %%CITATION = ARXIV:0708.1621;%%

\bibitem{Kalinowski:2012re} 
  M.~Kalinowski and M.~Wagner,
  %``Strange and charm meson masses from twisted mass lattice QCD,''
  PoS CONFINEMENT {\bf 10}, 303 (2012)
  [arXiv:1212.0403 [hep-lat]].
  %%CITATION = ARXIV:1212.0403;%%

\bibitem{Baron:2008xa}
  R.~Baron {\it et al.} [ETM Collaboration],
  % ``Status of ETMC simulations with $N_f = 2+1+1$ twisted mass fermions,''
  PoS {\bf LATTICE2008}, 094 (2008)
  [arXiv:0810.3807 [hep-lat]].
  %%CITATION = POSCI,LATTICE2008,094;%%

\bibitem{Baron:2009zq} 
  R.~Baron {\it et al.} [ETM Collaboration],
  % ``First results of ETMC simulations with $N_f = 2+1+1$ maximally twisted mass fermions,''
  PoS {\bf LATTICE2009}, 104 (2009)
  [arXiv:0911.5244 [hep-lat]].
  %%CITATION = ARXIV:0911.5244;%%

\bibitem{Baron:2010bv} 
  R.~Baron {\it et al.} [ETM Collaboration],
  % ``Light hadrons from lattice QCD with light $(u,d)$, strange and charm dynamical quarks,''
  JHEP {\bf 1006}, 111 (2010)
  [arXiv:1004.5284 [hep-lat]].
  %%CITATION = ARXIV:1004.5284;%%

\bibitem{Baron:2011sf} 
  R.~Baron {\it et al.} [ETM Collaboration],
  % ``Light hadrons from $N_f = 2+1+1$ dynamical twisted mass fermions,''
  PoS {\bf LATTICE2010}, 123 (2010)
  [arXiv:1101.0518 [hep-lat]].
  %%CITATION = ARXIV:1101.0518;%%

\bibitem{Baron:2010th} 
  R.~Baron {\it et al.} [ETM Collaboration],
  % ``Computing $K$ and $D$ meson masses with $N_f = 2+1+1$ twisted mass lattice QCD,''
  Comput.\ Phys.\ Commun.\ {\bf 182}, 299 (2011)
  [arXiv:1005.2042 [hep-lat]].
  %%CITATION = ARXIV:1005.2042;%%

\bibitem{Baron:2010vp} 
  R.~Baron {\it et al.} [ETM Collaboration],
  % ``Kaon and $D$ meson masses with $N_f = 2+1+1$ twisted mass lattice QCD,''
  PoS {\bf LATTICE2010}, 130 (2010)
  [arXiv:1009.2074 [hep-lat]].
  %%CITATION = ARXIV:1009.2074;%%

\bibitem{Jansen:2008si}
  K.~Jansen {\it et al.} [ETM Collaboration],
  % ``The static-light meson spectrum from twisted mass lattice QCD,''
  JHEP {\bf 0812}, 058 (2008)
  [arXiv:0810.1843 [hep-lat]].
  %%CITATION = ARXIV:0810.1843;%%

\bibitem{Boucaud:2008xu} 
  P.~Boucaud {\it et al.} [ETM Collaboration],
  %``Dynamical Twisted Mass Fermions with Light Quarks: Simulation and Analysis Details,''
  Comput.\ Phys.\ Commun.\  {\bf 179}, 695 (2008)
  [arXiv:0803.0224 [hep-lat]].
  %%CITATION = ARXIV:0803.0224;%%

\bibitem{Blossier:2009kd} 
  B.~Blossier, M.~Della Morte, G.~von Hippel, T.~Mendes and R.~Sommer,
  %``On the generalized eigenvalue method for energies and matrix elements in lattice field theory,''
  JHEP {\bf 0904}, 094 (2009)
  [arXiv:0902.1265 [hep-lat]].
  %%CITATION = ARXIV:0902.1265;%%

\bibitem{PDG}
  K. Nakamura {\it et al.} [Particle Data Group Collaboration],
  ``Review of particle physics,''
   J.\ Phys.\ G {\bf 37}, 075021 (2010) and 2011 partial update for the 2012 edition.

\bibitem{Ebert:2009ua} 
  D.~Ebert, R.~N.~Faustov and V.~O.~Galkin,
  %``Heavy-light meson spectroscopy and Regge trajectories in the relativistic quark model,''
  Eur.\ Phys.\ J.\ C {\bf 66}, 197 (2010)
  [arXiv:0910.5612 [hep-ph]].
  %%CITATION = ARXIV:0910.5612;%%

\bibitem{Alexandrou:2012rm} 
  C.~Alexandrou {\it et al.} [ETM Collaboration],
  %``Lattice investigation of the scalar mesons a_0(980) and \kappa\ using four-quark operators,''
  arXiv:1212.1418 [hep-lat].
  %%CITATION = ARXIV:1212.1418;%%

\bibitem{Blossier:2009vy} 
  B.~Blossier {\it et al.} [ETM Collaboration],
  %``Lattice calculation of the Isgur-Wise functions tau(1/2) and tau(3/2) with dynamical quarks,''
  JHEP {\bf 0906}, 022 (2009)
  [arXiv:0903.2298 [hep-lat]].
  %%CITATION = ARXIV:0903.2298;%%

\bibitem{Blossier:2009eg} 
  B.~Blossier {\it et al.} [ETM Collaboration],
  %``Dynamical lattice computation of the Isgur-Wise functions tau(1/2) and tau(3/2),''
  PoS {\bf LATTICE2009}, 253 (2009)
  [arXiv:0909.0858 [hep-lat]].

\end{thebibliography}
\end{document}